\documentclass{DISproc}

\begin{document}
%------------------------------------
\title{Insights into the Nucleon Spin from Lattice QCD}

%for single authors the superscripts are optional
\author{{\slshape Sara\ Collins$^1$ for the QCDSF Collaboration}\\[1ex]
${}^1$Institut f\"ur Theoretische Physik, Universit\"at Regensburg,\\
93040 Regensburg, Germany}

% please enter the contribution ID for the DOI
\contribID{123}

\doi  % if there is an online version we will register DOIs

\maketitle

\begin{abstract}
Flavour singlet contributions to the nucleon spin are elusive due to
the fact that they cannot be determined directly in experiment but
require extrapolations to the small x region. Direct calculations of
these contributions are possible using Lattice QCD, however, they pose
a significant computational challenge due to the presence of
disconnected quark line diagrams. We report on recent progress in
determining these sea quark contributions on the lattice.
\end{abstract}

\section{Introduction and Results}
The distribution of the spin of the proton among its
constituents has long been a topic of interest. The total spin
can be decomposed into the contribution from the quark spins,
$\Delta \Sigma$, the quark orbital angular momenta, $L_\psi$, and the
gluon total angular momentum $J_g$~\cite{Ji:1996ek},
\begin{eqnarray}
\frac{1}{2} &=&\frac{1}{2}\Delta\Sigma+L_\psi+J_g,\label{spin}
\end{eqnarray}
where $\Delta \Sigma = \Delta u+\Delta d+\Delta s$~(heavier quarks
are normally neglected). In this work, $\Delta q$~($q=u,d,s$) denotes
the combined spin contribution of the quark and the antiquark.  Using
Lattice QCD, one can determine the $\Delta q$ from first principles
through the axial-vector matrix element,
\begin{eqnarray}
\frac{1}{m_N}\langle N,s|\bar{q}\gamma_{\mu}\gamma_5\frac{1}{2} q|N,s\rangle & = & \frac{\Delta q}{2}\hspace{0.1cm}s_{\mu},
\end{eqnarray}
where $m_N$ is the mass of the nucleon with spin $s_\mu$~($s_\mu^2 =
-1)$.  Thus, one can construct the axial charges, $a_3 = \Delta u
-\Delta d = g_A$, $a_8 = \Delta u +\Delta d - 2\Delta s$ and $a_0(Q^2)
= \Delta u +\Delta d + \Delta s = \Delta \Sigma (Q^2)$.  $\Delta
\Sigma$ acquires a scale dependence, $Q^2$, due to the axial
anomaly. The axial-vector matrix element is related to the first
moment of the quark helicity distributions. The second moment, $\langle
x\rangle_\Delta$, and the second moment of the transverse helicity
distribution, $\langle x\rangle_\delta$, can also be calculated on the
lattice, higher moments are more challenging.  Furthermore, the total
angular momentum of quark $q$, $J_q=\frac{1}{2}\Delta q +L_q$~($\sum_q
J_q=\frac{1}{2}\Delta\Sigma+L_\psi$) can be obtained from the
generalised form factors, $A^q_{20}(Q^2)$ and $B^q_{20}(Q^2)$, which
parameterise the matrix element of the energy-momentum tensor for
momentum transfer, $Q$.  For a review of recent Lattice results of
$J_q$, $\langle x\rangle_\Delta$ and $\langle x\rangle_\delta$
see~\cite{Hagler:2009ni,Alexandrou:2011iu} and references therein.

In these proceedings we focus on $\Delta q$ and $\Delta s$ in
particular. Lattice results for $\Delta s$ have an important role to
play in constraining fits of polarised parton distribution
functions~(PDF).  The spin structure function of the proton and
neutron, $g^{n,p}_1(x, Q^2)$, is measured in deep inelastic
experiments. The first moment is related to the axial charges via the
operator product expansion. To leading twist:
\begin{eqnarray}
\Gamma^{p,n}_1(Q^2)  =  \int_0^1 \mathrm{d}x\, g_1^{p,n}(x,Q^2) & =& \frac{1}{36} \left[(a_8\pm 3a_3)C_{NS}+4a_0C_S\right]\label{int}
\end{eqnarray}
where $C_{S}$ and $C_{NS}$ are the singlet and non-singlet Wilson
coefficients, respectively. Model assumptions are made in order to
extrapolate $g^{n,p}_1(x, Q^2)$ from the minimum $x=0.02$ accessible
in experiment down to $x=0$. $a_3$ is known from neutron $\beta$-decay,
while, assuming $SU(3)$ flavour symmetry, $a_8$ can be obtained from
hyperon $\beta$-decays. Thus, in combination with
$\Gamma^{p,n}_1(Q^2)$, $a_0$ and the $\Delta q$ can be deduced. For
example, HERMES find $\Delta s (5 \mathrm{GeV}^2)=
\frac{1}{3}(a_0-a_8)=-0.085(13)(8)(9)$~\cite{Airapetian:2007mh}.
However, if the range of $x$ in the integral in Eq.~\ref{int} is
restricted to the experimental range, $x>0.02$, $\Delta s$ is
consistent with zero, indicating the large negative value arises from
model assumptions in the low $x$ region.

Semi-inclusive deep inelastic scattering~(SIDIS) experiments offer a
direct measurement of the $\Delta q$ using pion and kaon
beams. Results from COMPASS show the strangeness contribution is
consistent with zero down to
$x=0.004$~\cite{Alekseev:2010ub}. A naive extrapolation to
$x=0$ gives $\Delta s = -0.02(2)(2)$, while using the parameterisation
of De Florian et al.~(DSSV)~\cite{deFlorian:2008mr} gives $\Delta s =
-0.10(2)(2)$. Present measurements via SIDIS are limited by the
knowledge of the quark fragmentation functions, to which $\Delta s$ is
particularly sensitive. Another possibility of directly determining
$\Delta s$ combines $\nu p$, $\bar{\nu}p$ and parity-violating $ep$
elastic scattering data~\cite{Pate:2003rk}. Here, the MicroBooNE
experiment will enable errors to be significantly reduced.

Considering the lattice approach, simulations are performed at finite
volume~($V$) and lattice spacing~($a$) and typically with $u$ and $d$
quarks with unphysically heavy masses~($m_q$). Physical results are
recovered in the continuum~($a\to 0$) and infinite volume~($V\to
\infty$) limits at physical quark masses. The $u$, $d$ and $s$ quark
masses used in simulations are normally expressed in terms of the
pseudoscalar meson masses they correspond to~($M_{PS}^2\propto
m_q$). Developments in algorithmic techniques and computing power mean
typical simulations now involve lattices with $a\sim {0.05-0.1}$~fm,
$V\sim 2.5-6$~fm and $M_{PS}(u/d)\sim 200$~MeV.

For any lattice prediction, the size of the main systematic errors,
discretisation effects, finite volume and so on, must be investigated
thoroughly. However, the systematic uncertainties should always be
compared to the inherent statistical error, where for some quantities the
latter dominates. $\Delta s$ is one such example.  Evaluating $\langle
N|\bar{s}\gamma_{\mu}\gamma_5 s|N\rangle$ involves calculating a
disconnected quark line diagram~(for the strange quark). These types
of diagrams are computationally expensive to calculate as they involve
the quark propagator from all space-time lattice points to all points.
In the past these diagrams were often not calculated and differences
of quantities were quoted, for which the disconnected contribution
cancelled assuming isospin symmetry, for example, $g_A=\Delta u -
\Delta d$, $J_{u-d}$ and $\langle x\rangle_{\Delta u - \Delta
  d}$. However, methods have been developed which enable the
disconnected contributions to be calculated~\cite{Bali:2009hu}, in
particular, also $\Delta s$. Note that for $\Delta u$ and $\Delta d$ a
``connected'' quark line diagram must also be evaluated in addition to
the disconnected one.

In the following we present results for the quark spin contributions
to the proton generated on configurations with two degenerate flavours
of sea quarks\footnote{The number of sea quarks refers to the number
  of flavours of quark fields included in the Monte Carlo generation
  of an ensemble of representative quark and gluon field
  configurations.}, with $a\sim 0.072$~fm and $u/d$ quark masses given
by $M_{PS}=285$~MeV.  The quark action employed has leading order
discretisation effects of $O(a^2)$, which are not expected to be
significant for this value of the lattice spacing. Two lattice
volumes, with spatial dimensions of $2.3$~fm and $2.9$~fm, were used
and no significant finite volume effects were found. Although the $u$ and
$d$ quark masses are unphysically heavy, we varied the quark masses in
the range corresponding to $M_{PS}=285-720$~MeV and found no
significant change in the results. This suggests our results may also
apply to physical quark masses. Further details can be found
in~\cite{QCDSF:2011aa}.

\begin{table}[htb]
\centering
\begin{tabular}{ccc|c|cc}\toprule
$q$&$\Delta q^{\mathrm{lat}}_{\rm con}$&$\Delta q^{\mathrm{lat}}_{\rm dis}$&$\Delta q^{\mathrm{\overline{MS}}}(Q)$ & DSSV $x_{min}$ & DSSV $0$  \\\midrule
$u$        &~1.071(15)&-0.049(17)&~0.787(18)(2)& 0.793&0.814\\
$d$        &-0.369( 9)&-0.049(17)&-0.319(15)(1)& -0.416&-0.458\\
$s$& 0        &-0.027(12)& -0.020(10)(1) & -0.012&-0.114\\
$a_3$&~1.439(17)& 0        & ~1.105(13)(2) & 1.21&1.272\\
$a_8$      &~0.702(18)&-0.044(19)& ~0.507(20)(1)&0.401&0.583\\
$\Sigma$   &~0.702(18)&-0.124(44)& ~0.448(37)(2)&0.366&0.242\\\bottomrule
\end{tabular}
\caption{The bare~(lattice) connected, $\Delta q^{\mathrm{lat}}_{\rm
    con}$, and disconnected, $\Delta q^{\mathrm{lat}}_{\rm dis}$,
  quark spin contributions of the proton. Renormalisation to the
  $\mathrm{\overline{MS}}$ scheme is performed for
  $Q\sim\sqrt{7.4}$~GeV, to obtain $\Delta
  q^{\mathrm{\overline{MS}}}(Q)$. The first error is statistical, the
  second is from the renormalisation. A comparison is made with the
  results of DSSV global fits to DIS and SIDIS
  data~\protect\cite{deFlorian:2009vb}~(see this reference for the
  uncertainties on the values) for the cases where $x$ is restricted
  to $\left[x_{min}=0.001\to 1\right]$ and where $x=\left[0 \to
    1\right]$.}
\label{table:deltas}
\end{table}

\begin{figure}[htb]
\centerline{\includegraphics[width=.5\textwidth]{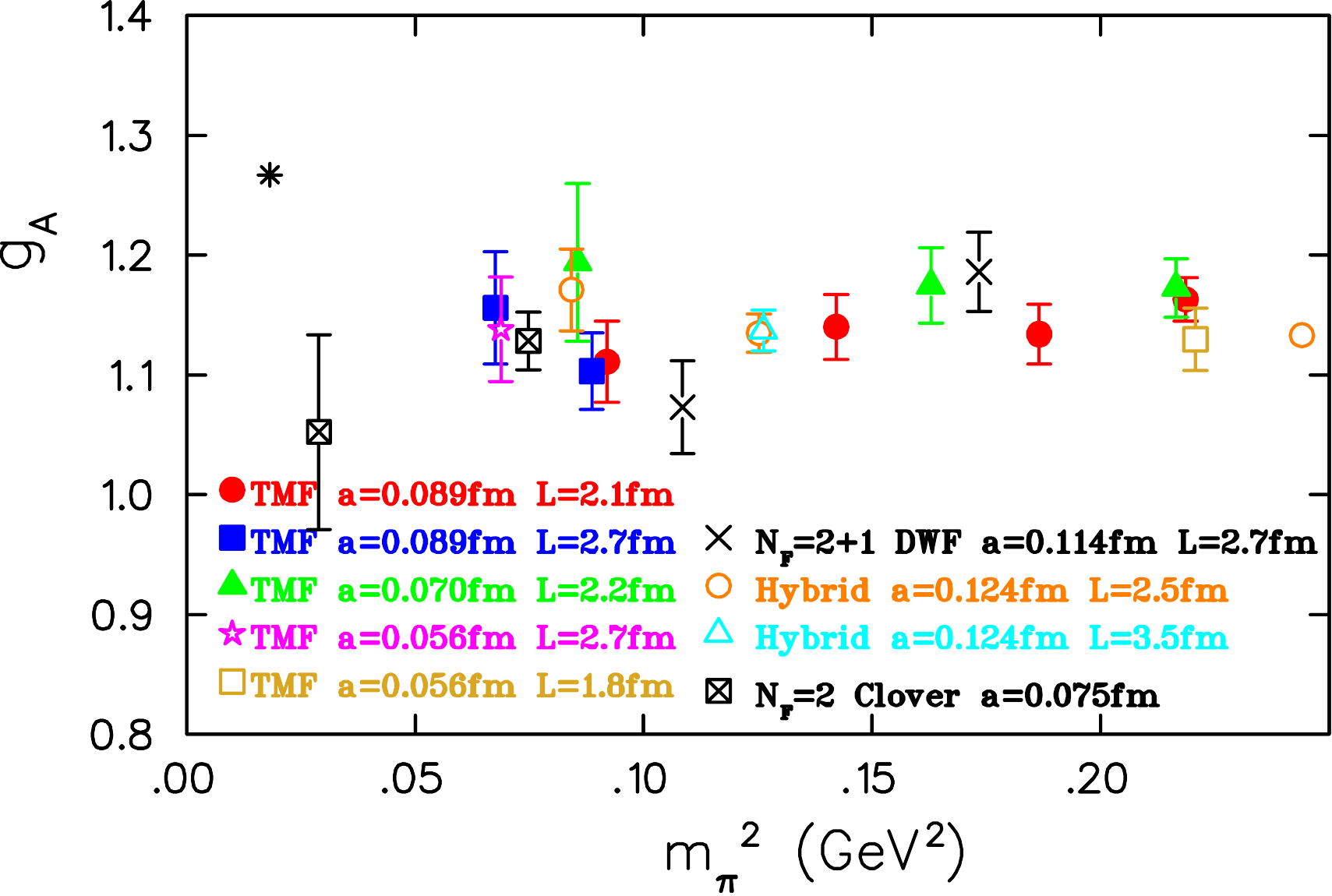}
\includegraphics[width=.5\textwidth]{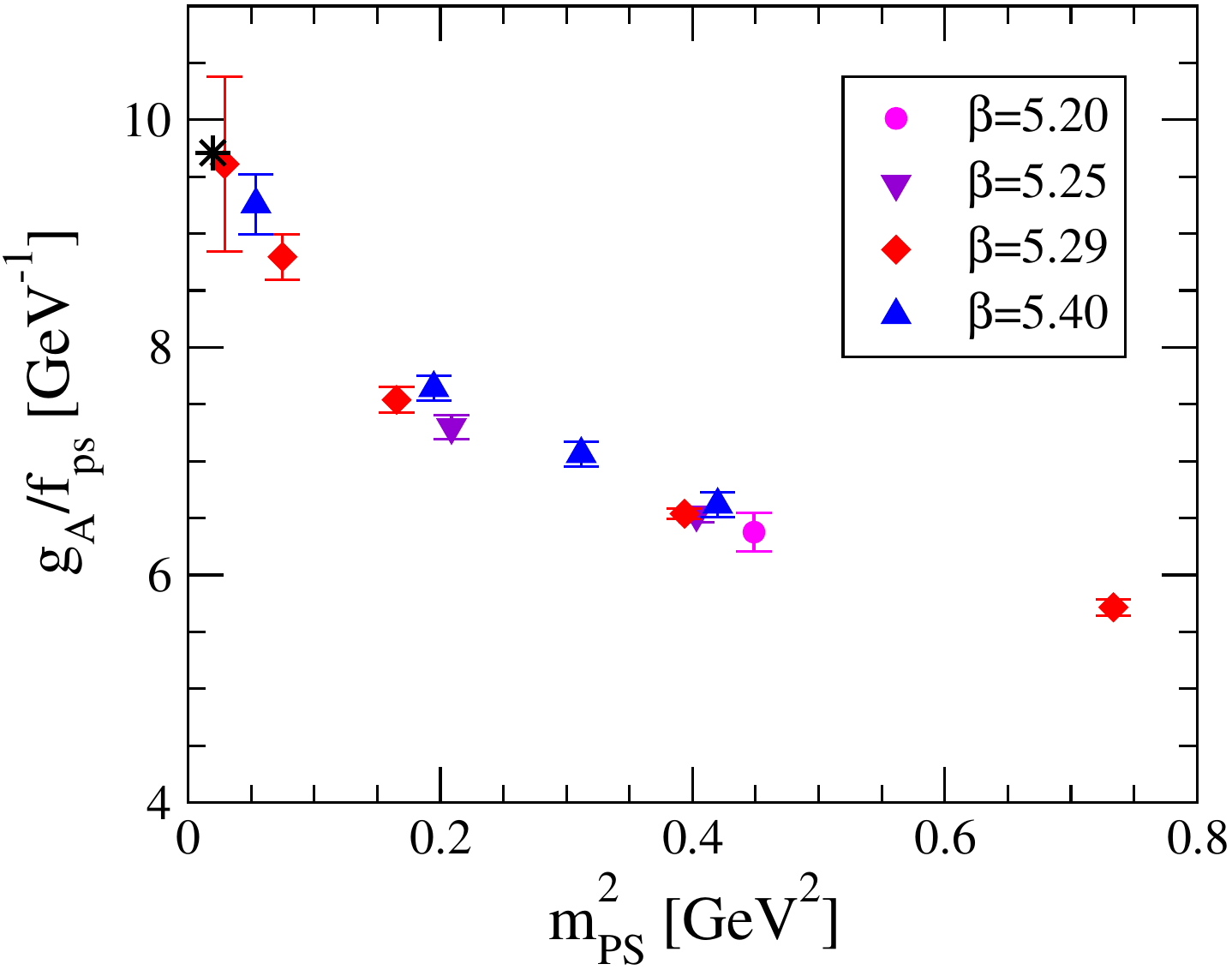}}
  \caption{(left) A summary plot of $a_3=g_A$ for different lattice
    simulations as a function of the $u/d$ quark mass, expressed in
    terms of
    $M_{PS}^2=m_\pi^2$~\protect\cite{Alexandrou:2011iu}. (right) The
    ratio of $g_A/f_\pi$ as a function of $M_{PS}^2$ for different
    lattice spacings~\protect\cite{Pleiter:2011gw}. $\beta=5.2-5.4$
    corresponds to $a\sim 0.083-0.060$~fm.}
  \label{Fig3}
\end{figure}

Table~\ref{table:deltas} displays the results for $\Delta s$ and the
axial charges and compares them to results from a DSSV global
analysis~\cite{deFlorian:2009vb}. We obtain a small, negative value
for $\Delta s$, consistent with the result obtained from a truncated
DSSV fit to experimental data. Notably, $a_3$ is significantly lower
than the experimental value of $a_3=g_A=1.2695(29)$. This is a general
feature of lattice calculations of $g_A$. A summary plot of recent
simulations taken from~\cite{Alexandrou:2011iu} is shown in
Fig.~\ref{Fig3}. Over the range of $u/d$ quark masses available the
results obtained using different lattice quark actions, volumes and
lattice spacings, are fairly constant and lie approximately $10\%$
below the experimental result. Simulations at smaller $u/d$ quark masses
are needed. Problems with excited state contaminations have also been
suggested as a possible source of the discrepancy, see, for example,
Ref.~\cite{Capitani:2012gj}. On the lattice an operator with the
quantum numbers of the proton, creates all states of
$J^P=\frac{1}{2}^+$ and one needs to ensure that the matrix element
for the ground state has been extracted. Nonetheless, the uncertainty
in the lattice determination of $g_A$ seems to be multiplicative, as
shown on the right in Fig.~\ref{Fig3}: the ratio of $g_A/f_\pi$ tends
to the experimental value.

Considering the underestimate of $g_A$, we add a $20\%$ uncertainty to
our results and obtain a final value of
\[
\Delta s = -0.020(10)(4).
\]
The first error is statistical, the second is due to the 
systematic uncertainty, which dominates. Other groups using
similar methods, albeit at heavier quark masses and without
renormalisation, obtain consistent results, see~\cite{Babich:2010at}
and~\cite{Engelhardt:2010zr}.  This is in contradiction to earlier
exploratory work by, for example, the Kentucky
group~\cite{Dong:1995rx}. Note that this group have recently also
calculated flavour singlet contributions to $J_q$~\cite{Liu:2012nz}.

\section{Acknowledgements}

This work is supported by the EU ITN STRONGnet  and the
DFG SFB/TRR 55. S.C. acknowledges support from the
Claussen-Simon-Foundation (Stifterverband f\"ur die Deutsche
Wissenschaft).

{\raggedright
\begin{footnotesize}
 \bibliographystyle{DISproc}
 \bibliography{collins_sara.bib}
\end{footnotesize}
}
\end{document}